\documentclass[aps,prl,twocolumn,showkeys,showpacs]{revtex4}

\usepackage{amsmath}
\usepackage{amssymb}
\usepackage{dcolumn}

\begin{document}
 
\title{Quantum teleportation of an arbitrary two qubit state and its relation to multipartite entanglement}

\author{Gustavo Rigolin}
\affiliation{Departamento de Raios C\'osmicos e Cronologia, Instituto de F\'{\i}sica Gleb Wataghin, Universidade Estadual de Campinas, C.P. 6165, cep 13084-971, Campinas, S\~ao Paulo, Brazil}

\begin{abstract}
We explicitly show a protocol in which an arbitrary two qubit $|\phi \rangle$ $=$ $a|00\rangle$ $+$ $b|01\rangle$ $+$ $c|10\rangle$ + $d|11\rangle$ is faithfully and deterministically teleported from Alice to Bob. We construct the $16$ orthogonal generalized Bell states which can be used to teleport the two qubits. The local operations Bob must perform on his qubits in order to recover the teleported state is also constructed. They are restricted only to single qubit gates. This means that a CNOT gate is not necessary to complete the protocol. A generalization where $N$ qubits is teleported is also shown. We define a generalized magic basis, which possesses interesting properties. These properties help us to suggest a generalized concurrence from which we construct a new measure of entanglement that has a clear physical interpretation: A multipartite state has maximum entanglement if it is a genuine quantum teleportation channel. 
\end{abstract}



\maketitle

In $1993$ Bennett \textit{et al} \cite{bennett1} discovered one of the most astonishing features of Quantum Mechanics: The Quantum Teleportation. Using what they called an Einstein-Podolsky-Rosen Channel ($|\phi^{+}\rangle$ $=$ $1/\sqrt{2}$ ($|00\rangle$ $+$ $|11\rangle$), for example) they showed that it is possible to transmit a one qubit state $a|0\rangle + b|1\rangle$ from one location (Alice) to another (Bob) sending $2$ bits of classical information. Four years later, this protocol was experimentally demonstrated \cite{zeilinger}.

The next natural question was whether it was possible to teleport not just a single qubit, but rather $N$ qubits. Lee \textit{et al} \cite{lee}, in a very interesting work, showed that it was possible to teleport a two qubit state $|\Phi \rangle$ $=$ $a|00\rangle$ $+$ $b|01\rangle$ $+$ $c|10\rangle$ + $d|11\rangle$ from Alice to Bob using a four entangled state and sending him $4$ bits of classical information. However they did not explicitly constructed the protocol and did not provided a generalization to $N$ qubits. 

Here we explicitly construct this protocol and present a generalization to $N$ qubits. We create a new set of $16$ generalized Bell states to implement the teleportation. These states slightly differ from the generalized Bell states used in a probabilistic two qubit teleportation protocol by Gao \textit{et al} \cite{gao}. We also show that the unitary operations Bob must apply on his qubits to faithfully obtain the teleported two qubits are restricted to single qubit gates, i. e., gates such as CNOT are not necessary to accomplish the protocol.

The motivation for defining these new set of $16$ generalized Bell states lies in the fact that using them, we can easily construct a ``magic basis''. We show that this magic basis possesses the same interesting properties of the original one defined in Ref. \cite{bennett2}. With the aid of the magic basis we generalize Wootters concurrence \cite{wootters} and define the Entanglement of Teleportation ($E_{T}$), which has a simple physical interpretation in terms of the efficiency an entangled $2N$ qubit state has to teleport an unknown $N$ qubit state. 

In the original proposal \cite{bennett1} the teleportation of a single qubit $|\phi \rangle$ $=$ $a|0\rangle + b|1\rangle$ is executed as follows. Alice and Bob initially share a maximally two qubit entangled state: $|\Psi^{-}\rangle$ $=$ $1/\sqrt{2}$ ($|01\rangle$ $-$ $|10\rangle)$. The joint system (qubit to be teleported plus EPR state) before Alice's measurement can be written as
\begin{eqnarray}
|\Phi\rangle & = & |\phi\rangle \otimes |\Psi^{-}\rangle \nonumber \\
|\Phi\rangle & = & \frac{a}{\sqrt{2}}\left( |001\rangle - |010\rangle \right) + \frac{b}{\sqrt{2}}\left( |101\rangle - |110\rangle \right), \label{prima}
\end{eqnarray}
where we use the convention that the first two qubits belong to Alice and the third one belongs to Bob ($|AAB\rangle$, $A \rightarrow$ Alice and $B \rightarrow$ Bob). Rewriting Eq.~(\ref{prima}) in terms of the four Bell states $|\Phi^{\pm}\rangle$ $=$ $(1/\sqrt{2})$ $(|00\rangle$ $\pm$ $|11\rangle)$ and $|\Psi^{\pm}\rangle$ $=$ $(1/\sqrt{2})$ $(|01\rangle$ $\pm$ $|10\rangle)$ we get
\begin{eqnarray}
|\Phi\rangle & = & \frac{1}{2}\left\{ |\Psi^{-}\rangle (-a|0\rangle -b|1\rangle ) + |\Psi^{+}\rangle (-a|0\rangle + b|1\rangle )\right. \nonumber \\
& & \left. +  |\Phi^{-}\rangle (a|1\rangle + b|0\rangle ) +  |\Phi^{+}\rangle (a|1\rangle - b|0\rangle ) \right\}. 
\end{eqnarray}     
Alice now makes a Bell measurement and communicates classically the result to Bob. With this $2$ bit information at hand, Bob applies the appropriate unitary operation to obtain the state $|\phi\rangle$. See Table \ref{tabela1}.
\begin{table}[!ht]
\vspace{-.35cm}
\caption{\label{tabela1} The unitary transformations Bob must perform on his qubit, conditioned to Alice's measurement result, to complete the teleportation protocol. $I$ is the identity operator and $\sigma$ are the usual Pauli matrices.}
\begin{ruledtabular}
\begin{tabular}{lll}
 Alice's result & Bob's operation & Bob's qubit \\ \hline
$|\Psi^{-}\rangle$ & $I$ & $I (-a|0\rangle -b|1\rangle) \propto |\phi\rangle$ \\
$|\Psi^{+} \rangle$ & $\sigma^{z}$ & $\sigma^{z}(-a|0\rangle + b|1\rangle ) \propto|\phi\rangle$  \\ 
$|\Phi^{-} \rangle$ & $\sigma^{x}$ & $\sigma^{x} (a|1\rangle + b|0\rangle ) \propto|\phi\rangle $ \\
$|\Phi^{+} \rangle$ & $\sigma^{z}\sigma^{x}$ & $\sigma^{z}\sigma^{x}(a|1\rangle - b|0\rangle ) \propto |\phi\rangle $ 
\end{tabular}
\end{ruledtabular}
\vspace{-.25cm}
\end{table}

Let us now explicitly present the protocol to teleport arbitrary two qubits. The state Alice wants to teleport is written as 
\begin{equation}
|\phi \rangle = a|00\rangle + b|01\rangle + c|10\rangle + d|11\rangle,
\end{equation} 
where $a, b, c$, and $d$ are complex coefficients and we assume $|\phi \rangle$ to be normalized. We now define the $16$ Generalized Bell states \cite{rigolin}, or G-states for simplicity. We divide them in four groups.

\noindent Group $1$:
\vspace{-.35cm}
\begin{eqnarray}
|g_{1}\rangle & = & \frac{1}{2}\left( |0000\rangle +|0101\rangle + |1010\rangle +|1111\rangle \right), \label{g1} \\
|g_{2}\rangle & = & \frac{1}{2}\left( |0000\rangle +|0101\rangle - |1010\rangle - |1111\rangle \right), \\
|g_{3}\rangle & = & \frac{1}{2}\left( |0000\rangle -|0101\rangle + |1010\rangle - |1111\rangle \right), \\
|g_{4}\rangle & = & \frac{1}{2}\left( |0000\rangle -|0101\rangle - |1010\rangle + |1111\rangle \right).
\end{eqnarray}
Group $2$:
\vspace{-.35cm}
\begin{eqnarray}
|g_{5}\rangle & = & \frac{1}{2}\left( |0001\rangle +|0100\rangle + |1011\rangle + |1110\rangle \right), \\
|g_{6}\rangle & = & \frac{1}{2}\left( |0001\rangle +|0100\rangle - |1011\rangle - |1110\rangle \right), \\
|g_{7}\rangle & = & \frac{1}{2}\left( |0001\rangle -|0100\rangle + |1011\rangle - |1110\rangle \right), \\
|g_{8}\rangle & = & \frac{1}{2}\left( |0001\rangle -|0100\rangle - |1011\rangle + |1110\rangle \right).
\end{eqnarray}
Group $3$:
\vspace{-.35cm}
\begin{eqnarray}
|g_{9}\rangle & = & \frac{1}{2}\left( |0010\rangle +|0111\rangle + |1000\rangle + |1101\rangle \right), \\
|g_{10}\rangle & = & \frac{1}{2}\left( |0010\rangle +|0111\rangle - |1000\rangle - |1101\rangle \right), \\
|g_{11}\rangle & = & \frac{1}{2}\left( |0010\rangle -|0111\rangle + |1000\rangle - |1101\rangle \right), \\
|g_{12}\rangle & = & \frac{1}{2}\left( |0010\rangle -|0111\rangle - |1000\rangle + |1101\rangle \right).
\end{eqnarray}
Group $4$:
\vspace{-.35cm}
\begin{eqnarray}
|g_{13}\rangle & = & \frac{1}{2}\left( |0011\rangle +|0110\rangle + |1001\rangle + |1100\rangle \right), \\
|g_{14}\rangle & = & \frac{1}{2}\left( |0011\rangle +|0110\rangle - |1001\rangle - |1100\rangle \right), \\
|g_{15}\rangle & = & \frac{1}{2}\left( |0011\rangle -|0110\rangle + |1001\rangle - |1100\rangle \right), \\
|g_{16}\rangle & = & \frac{1}{2}\left( |0011\rangle -|0110\rangle - |1001\rangle + |1100\rangle \right). \label{g16}
\end{eqnarray}
These states form an orthonormal basis, $\sum_{j=1}^{16}|g_{j}\rangle\langle g_{j}|$ $=$ $I$ and $\langle g_{j}|g_{k} \rangle$ $=$ $\delta_{jk}$, which we call Generalized Bell basis \cite{rigolin}, or G-basis for brevity.

Alice and Bob must share one of the $16$ G-states, where we assume that the first two qubits are with Alice and the last two with Bob. These last two qubits with Bob will be used to ``receive'' the teleported state. For definiteness we assume Alice and Bob share the $|g_{1}\rangle$ state. Hence, the initial joint state is
\begin{eqnarray}
|\Phi \rangle & = &|\phi\rangle \otimes |g_{1}\rangle \nonumber \\
 & = & \frac{a}{2} \left\{ |000000\rangle + |000101\rangle +  |001010\rangle +  |001111\rangle \right\} \nonumber \\
&  & + \frac{b}{2} \left\{ |010000\rangle + |010101\rangle +  |011010\rangle +  |011111\rangle \right\} \nonumber \\
&  & + \frac{c}{2} \left\{ |000000\rangle + |100101\rangle +  |101010\rangle +  |101111\rangle \right\} \nonumber \\
&  & + \frac{d}{2} \left\{ |110000\rangle + |110101\rangle +  |111010\rangle +  |111111\rangle \right\} \nonumber \\
\label{20}
\end{eqnarray} 
Here the first four qubits belong to Alice and the last two belong to Bob ($|AAAABB\rangle$, $A \rightarrow$ Alice and $B \rightarrow$ Bob). Using Eqs.~(\ref{g1}-\ref{g16}) we can write Eq.~(\ref{20}) as
\begin{equation}
|\Phi \rangle = \frac{1}{4}\sum_{j=1}^{16}|g_{j}\rangle_{A}|\phi_{j}\rangle_{B}, \label{geral1} 
\end{equation}
where $A$ and $B$ is written to emphasize which qubits are with Alice and Bob and the states $|\phi_{j}\rangle$ are defined in Table \ref{tabela2}.
\begin{table}[!ht]
\vspace{-.35cm}
\caption{\label{tabela2} The first column shows the $|\phi_{j}\rangle$ states. The third column depicts the local unitary operations Bob must perform in his qubits, conditioned on Alice's measurement given in the second column, to finish the teleportation protocol.}
\begin{ruledtabular}
\begin{tabular}{lll}
$|\phi_{j}\rangle$ & Alice's result & Bob's operation\\ \hline
$|\phi_{1}\rangle = |\phi\rangle$ & $|g_{1} \rangle$ & $I$ \\ 
$|\phi_{2}\rangle = \sigma_{1}^{z} |\phi\rangle$ & $|g_{2} \rangle$ & $\sigma_{1}^{z}$ \\
$|\phi_{3}\rangle = \sigma_{2}^{z} |\phi\rangle$ & $|g_{3} \rangle$ & $\sigma_{2}^{z}$ \\
$|\phi_{4}\rangle =  \sigma_{1}^{z}\sigma_{2}^{z} |\phi\rangle$ & $|g_{4} \rangle$ &  $\sigma_{2}^{z} \sigma_{1}^{z}$ \\
$|\phi_{5}\rangle = \sigma_{2}^{x} |\phi\rangle$ & $|g_{5} \rangle$ &  $\sigma_{2}^{x}$ \\
$|\phi_{6}\rangle =  \sigma_{2}^{x} \sigma_{1}^{z} |\phi\rangle$ & $|g_{6} \rangle$ &  $\sigma_{1}^{z} \sigma_{2}^{x}$\\
$|\phi_{7}\rangle =  \sigma_{2}^{x}\sigma_{2}^{z} |\phi\rangle$ & $|g_{7} \rangle$ &  $\sigma_{2}^{z} \sigma_{2}^{x}$ \\
$|\phi_{8}\rangle =  \sigma_{2}^{x}\sigma_{2}^{z}\sigma_{1}^{z} |\phi\rangle$ &$|g_{8} \rangle$ & $\sigma_{1}^{z} \sigma_{2}^{z} \sigma_{2}^{x}$ \\
$|\phi_{9}\rangle = \sigma_{1}^{x} |\phi\rangle$ & $|g_{9} \rangle$ & $\sigma_{1}^{x}$ \\
$|\phi_{10}\rangle =  \sigma_{1}^{x} \sigma_{1}^{z} |\phi\rangle$ & $|g_{10} \rangle$ &  $\sigma_{1}^{z} \sigma_{1}^{x}$ \\
$|\phi_{11}\rangle =  \sigma_{1}^{x}\sigma_{2}^{z} |\phi\rangle$ & $|g_{11} \rangle$ & $\sigma_{2}^{z} \sigma_{1}^{x}$ \\
$|\phi_{12}\rangle =  \sigma_{1}^{x}\sigma_{1}^{z}\sigma_{2}^{z} |\phi\rangle$ & $|g_{12} \rangle$ & $\sigma_{2}^{z} \sigma_{1}^{z} \sigma_{1}^{x}$ \\
$|\phi_{13}\rangle =  \sigma_{1}^{x} \sigma_{2}^{x} |\phi\rangle$ & $|g_{13} \rangle$ & $\sigma_{2}^{x} \sigma_{1}^{x}$ \\
$|\phi_{14}\rangle =  \sigma_{1}^{x}\sigma_{2}^{x}\sigma_{1}^{z} |\phi\rangle$ & $|g_{14} \rangle$ & $\sigma_{1}^{z} \sigma_{2}^{x}\sigma_{1}^{x}$ \\
$|\phi_{15}\rangle =  \sigma_{1}^{x}\sigma_{2}^{x}\sigma_{2}^{z} |\phi\rangle$ & $|g_{15} \rangle$ & $\sigma_{2}^{z} \sigma_{2}^{x} \sigma_{1}^{x}$ \\
$|\phi_{16}\rangle =  \sigma_{1}^{x}\sigma_{2}^{x}\sigma_{1}^{z}\sigma_{2}^{z} |\phi\rangle$ & $|g_{16} \rangle$ & $\sigma_{2}^{z} \sigma_{1}^{z} \sigma_{2}^{x}\sigma_{1}^{x}$ 
\end{tabular}
\end{ruledtabular}
\vspace{-.25cm}
\end{table}

Alice now makes a Generalized Bell measurement (G-measurement) obtaining with equal probabilities one of the $16$ G-states. By G-measurement we mean that Alice makes a joint measurement in the two qubits she wants to teleport plus the two qubits from the G-state shared with Bob. Then she sends Bob a classical message of $4$ bits to inform which G-state she has measured. With this information Bob knows what unitary operation (Table \ref{tabela2}) he must apply on his two qubits to faithfully recover the teleported state. After applying the unitary operation the protocol is finished and Alice has succeeded in teleporting her arbitrary two qubit state.

Paying attention to the $16$ unitary operations which Bob should apply on his two qubits, we see that they all can be written as $U = U_{5} \otimes U_{6}$. Here $5$ and $6$ refer to the fifth and sixth qubit respectively, i. e., Bob's two qubits. This means that we only need single qubit gates to implement all the $16$ unitary operations. More elaborated two qubit gates, such as the CNOT gate, is not necessary. This fact possibly simplifies future experimental realizations of this protocol. It is worth mentioning that we can construct other sets of $16$ orthogonal states which faithfully teleport any two qubits, but now depending on Alice's outcome, Bob will need to implement a CNOT gate to complete the protocol. 

If we use quantum channels with generalized GHZ states \cite{rigolin}, $|GHZ\rangle$ $=$ $1/\sqrt{2}$$(|0000\rangle$ $+$ $|1111\rangle)$, the protocol does not work. It is impossible to teleport an arbitrary two qubit state using a GHZ state. Only special classes of two qubit states such as $b|01\rangle+c|10\rangle$ can be teleported \cite{lee2}. 

The technical difficulties which we need to circumvent to experimentally test this protocol are not trivial. Three benchmarks must be reached to implement this protocol. First, Alice and Bob should have a source of four maximally entangled states (G-states). Second, Bob should be able to implement the $16$ local unitary operations, and Alice should somehow make a G-measurement. The first two benchmarks is very close \cite{4photons,5photons}, but we still need an efficient way to discriminate the $16$ G-states.

The previous protocol can be generalized to teleport a $N$ qubit state. For this purpose Alice needs to share a $2 N$ entangled G-state with Bob. Then she makes a $2 N$ joint G-measurement with the $N$ qubit to be teleported and half of the shared G-state and sends a $2N$ bit classical message to Bob informing the measurement outcome. Bob finishes the protocol performing at most $2N$ single qubit gates to obtain the teleported state. The number of unitary operations Bob needs to apply in his $N$ qubits is conditioned on Alice's outcome. The $N$ qubit teleportation protocol can be rigorously constructed as follows: (1) Generate the seed G-state $|s_{0}\rangle$ $=$ $(2^{-N/2})$ $\sum_{j=0}^{M}$ $|x_{j}\rangle_{A}$ $|x_{j}\rangle_{B}$, where $M=2^{N}-1$ and $x_{j}$ is the binary representation of the number $j$. In the two qubit teleportation protocol, $x_{0}=00$, $x_{1}=01$, $x_{2}=10$, and $x_{3}=11$. Zeros must be added to make all $x_{j}$ with the same amount of bits ($N$ bits). This G-state is our quantum channel and it is composed of $2 N$ qubits. (2) Using the seed G-state it is possible to obtain all the G-states locally operating on its first $N$ qubits, $|s_{j}\rangle = \bigotimes_{k=1}^{N}(\sigma_{k}^{z})^{j_{2k-1}}(\sigma_{k}^{x})^{j_{2k}}|s_{0}\rangle$. Now $j_{k}$ represents the $k$-th bit (from right to left) of the number $0 \leq j \leq 2^{2N} - 1$, which is written in binary notation and again zeros should be added to leave all $j$'s with the same number of bits ($2N$ bits). The subindex $k$ indicate on which qubit the Pauli matrices $\sigma^{x}$ and $\sigma^{z}$ should operate. For the two qubit protocol shown above, $|s_{0}\rangle = |g_{1}\rangle$, $|s_{1}\rangle = |g_{2}\rangle$, $|s_{2}\rangle = |g_{9}\rangle$, $|s_{3}\rangle = |g_{10}\rangle$, and so forth. (3) Alice makes a joint G-measurement with the $N$ qubits to be teleported and with her $N$ qubits of the shared G-state. She then sends Bob a $2N$ bit classical message informing Bob the measurement outcome. (4) With this information Bob applies the corresponding unitary operation on his $N$ qubits. These operations are given by $U_{j} = \bigotimes_{k=1}^{N}(\sigma_{k}^{z})^{j_{2k-1}}(\sigma_{k}^{x})^{j_{2k}}$ and for the two qubit teleportation protocol they are shown in Table \ref{tabela2}.   

We now define the generalized magic basis ($|e_{j}\rangle$) and an auxiliary basis ($|f_{j}\rangle$), which help us in the calculations that follows. See Table \ref{tabela3}.
\begin{table}[!ht]
\vspace{-.35cm}
\caption{\label{tabela3} The first column shows the magic states. The second column represents the corresponding G-states, and  the the third column the F-states. The contents of the rows should be read as $a$ $=$ $b$ $=$ $c$.}
\begin{ruledtabular}
\begin{tabular}{lll}
Magic states & G-states & F-states\\ \hline
$|e_{1} \rangle$ & $|g_{1} \rangle$ & $|f_{1} \rangle$ \\ 
$|e_{2} \rangle$ & $\mathrm{i}|g_{2} \rangle$ & $\mathrm{i} |f_{2} \rangle$ \\
$|e_{3} \rangle$ & $|g_{4} \rangle$ & $|f_{3} \rangle$ \\
$|e_{4} \rangle$ & $\mathrm{i}|g_{3} \rangle$ & $\mathrm{i}|f_{4} \rangle$ \\
$|e_{5} \rangle$ & $|g_{6} \rangle$ & $|f_{5} \rangle$ \\ 
$|e_{6} \rangle$ & $\mathrm{i}|g_{5} \rangle$ & $\mathrm{i} |f_{6} \rangle$ \\
$|e_{7} \rangle$ & $|g_{7} \rangle$ & $|f_{7} \rangle$ \\
$|e_{8} \rangle$ & $\mathrm{i}|g_{8} \rangle$ & $\mathrm{i}|f_{8} \rangle$ \\
$|e_{9} \rangle$ & $|g_{10} \rangle$ & $|f_{9} \rangle$ \\ 
$|e_{10} \rangle$ & $\mathrm{i}|g_{9} \rangle$ & $\mathrm{i} |f_{10} \rangle$ \\
$|e_{11} \rangle$ & $|g_{11} \rangle$ & $|f_{11} \rangle$ \\
$|e_{12} \rangle$ & $\mathrm{i}|g_{12} \rangle$ & $\mathrm{i}|f_{12} \rangle$ \\
$|e_{13} \rangle$ & $|g_{13} \rangle$ & $|f_{13} \rangle$ \\ 
$|e_{14} \rangle$ & $\mathrm{i}|g_{14} \rangle$ & $\mathrm{i} |f_{14} \rangle$ \\
$|e_{15} \rangle$ & $|g_{16} \rangle$ & $|f_{15} \rangle$ \\
$|e_{16} \rangle$ & $\mathrm{i}|g_{15} \rangle$ & $\mathrm{i}|f_{16} \rangle$
\end{tabular}
\end{ruledtabular}
\vspace{-.25cm}
\end{table}        
In terms of the F-states, a general pure two qubit state can be written as $|\Psi \rangle$ $=$ $\sum_{j=1}^{16}$ $\alpha_{j}|f_{j}\rangle$. From this state we define $|\tilde{\Psi} \rangle$ $=$ $\sum_{j=1}^{16}$ $\alpha_{j}^{*}|f_{j}\rangle$, where $\alpha_{j}^{*}$ means the complex conjugated of $\alpha_{j}$. With the previous definitions, we can present the generalized concurrence \cite{wong} as
\begin{equation}
C(\Psi) = \left| \langle \tilde{\Psi} | \sigma_{y}^{\otimes 4} | \Psi \rangle \right|, \label{gconc1}
\end{equation}
where $\sigma_{y}^{\otimes 4} = \sigma_{1}^{y}\sigma_{2}^{y}\sigma_{3}^{y}\sigma_{4}^{y}$. Since $\sigma_{y}^{\otimes 4}$ $| \Psi \rangle$ $=$ $\sum_{j=1}^{16}$ $(-1)^{j+1}$ $\alpha_{j}$ $|f_{j}\rangle$, 
\begin{equation}
C(\Psi) = \left| \sum_{j=1}^{16}(-1)^{j+1}\alpha^{2}_{j}\right|. 
\end{equation}
But in the magic basis $|\Psi \rangle$ $=$ $\sum_{j=1}^{16}$ $\beta_{j}|e_{j}\rangle$ and using the relations between the F-states and the magic states in Table \ref{tabela3} we can show that $\alpha_{j}=\mathrm{i}^{(j+1)\oplus 2}\beta_{j}$, where $a \oplus 2 = 0$ if $a$ is even and $1$ if $a$ is odd. This implies that
\begin{equation}
C(\Psi) = \left| \sum_{j=1}^{16}\beta^{2}_{j}\right|. \label{gconc2}
\end{equation}    
Using Eq.~(\ref{gconc2}) we can show that Eq.~(\ref{gconc1}) satisfies the same properties of the original Wootters concurrence \cite{wootters}. (1) If all $\beta_{j}$ are real then $C = 1$; (2) $0 \leq C \leq 1$; (3) in the magic basis, $C = |\langle \tilde{\Psi}|\Psi \rangle|$; (4) every two qubit state with $C=1$ can be written, up to a global phase, as a combination of the magic states with real coefficients; and (5) all the magic states have $C = 1$. Noting that $\sigma_{y}^{\otimes 4}$ is the spin flip operator, we can show that if $|\Psi\rangle$ is separable then $C = 0$.  

In order to quantify the usefulness a four qubit state has to teleport two qubits we introduce the Entanglement of Teleportation ($E_{T}$):
\begin{equation}
E_{T}(\Psi) = \frac{1}{16}\sum_{j=1}^{L}C(\Psi_{j}), \label{ent}
\end{equation} 
where $|\Psi_{j}\rangle$ are all the $L \leq 16$ orthogonal states which can be obtained from $|\Psi\rangle$ using the $16$ unitary operations $U_{j}$ listed in the third column of Table \ref{tabela2}. It should be noted that Eqs.~(\ref{gconc1}) and (\ref{ent}) can easily be extended to a $2N$ qubit state:
\begin{eqnarray}
C(\Psi) & = & \left| \langle \tilde{\Psi} | \sigma_{y}^{\otimes 2N} | \Psi \rangle \right|, \\
E_{T}(\Psi) & = & \frac{1}{2^{2N}}\sum_{j=1}^{L}C(\Psi_{j}),
\end{eqnarray}
where now $|\Psi_{j}\rangle$ are the $L \leq 2^{2N}$ orthogonal states obtained from $|\Psi\rangle$ using the $2^{2N}$ unitary operations $U_{j}= \bigotimes_{k=1}^{N}(\sigma_{k}^{z})^{j_{2k-1}}(\sigma_{k}^{x})^{j_{2k}}$. $E_{T}$ satisfies some interesting properties. (1) It can discriminate the generalized W, GHZ, and G-states, i. e. $E_{T}(W) < E_{T}(GHZ) < E_{T}(g_{j})$; (2) It can be seen as a measure of the efficiency  $2N$ qubits can teleport $N$ qubits; (3) All G-states have $E_{T}=1$; (4) All separable states have $E_{T} = 0$.

In order to illustrate the above properties, let us return to the four qubit case. Consider the G-state $|g_{1}\rangle$. Applying the $U_{j}$ operations we obtain $16$ orthogonal states (the G-basis) with $C=1$. Therefore, $E_{T}=1$. Now let us study the generalized GHZ state $|GHZ^{+}\rangle = 1/\sqrt{2}(|0000\rangle + |1111\rangle)$. The $16$ unitary operations $U_{j}$ produce only eight orthogonal states with $C = 1$,
\begin{eqnarray}
|GHZ^{\pm}\rangle & = & \frac{1}{\sqrt{2}}(|0000\rangle \pm |1111\rangle), \\
|G^{\pm}\rangle & = & \frac{1}{\sqrt{2}}(|0100\rangle \pm |1011\rangle), \\
|H^{\pm}\rangle & = & \frac{1}{\sqrt{2}}(|1000\rangle \pm |0111\rangle), \\
|Z^{\pm}\rangle & = & \frac{1}{\sqrt{2}}(|1100\rangle \pm |0011\rangle),
\end{eqnarray} 
which implies that $E_{T} = 1/2$. Repeating the same procedure with the generalized W state $|W\rangle = (1/2)(|0001\rangle+|0010\rangle+|0100\rangle+|1000\rangle)$ we obtain eight orthogonal states with $C = 0$, i. e., $E_{T} = 0$. We can get a physical picture of the meaning of these values for $E_{T}$ noting that using G-states ($E_{T}=1$) we can deterministically teleport any two qubits, using GHZ states ($E_{T}=1/2$) only especial classes of two qubits can be teleported \cite{lee2} and for $W$ states even these special classes cannot be deterministically teleported. And more, we can show that the GHZ states can deterministically teleport one qubit but the W states accomplish the same task only probabilistically.
  
We have explicitly shown a teleportation protocol which allows Alice to faithfully teleport an arbitrary two qubit state to Bob sending him a $4$ bit classical message. We generalized this protocol for the $N$ qubit case, where Alice can teleport $N$ qubits to Bob sending him a $2N$ bit classical message. They must also share a $2N$ entangled state, which is lost at the end of the protocol. We also presented a way to quantify the efficiency a $2N$ entangled state can faithfully teleport $N$ qubits, defining the Entanglement of Teleportation ($E_{T}$). $E_{T}$ is able to quantify the generalized Bell states, the GHZ states, and W states in terms of their teleportation capacities. We think that this new characterization of entanglement will be useful in the study of multipartite states, since it was shown that it has a simple physical interpretation, is operational,  and is easily generalized to states with an even number of qubits.   

\begin{acknowledgments}
The author thanks FAPESP for funding and Dr. Fernando da Rocha Vaz Bandeira de Melo for initiating a discussion which culminated in this work. 
\end{acknowledgments}

\textit{Note Added at Proof:} After completion of this work we became aware of an interesting work done by M. Ikram \textit{et al.} \cite{ikram} showing how to experimentally teleport $N$ qubits using $N$ high-Q cavities.

\end{document}